\documentclass[%
 reprint,
 amsmath,amssymb,
 aps,
]{revtex4-2}

\usepackage{graphicx}% Include figure files
\usepackage{dcolumn}% Align table columns on decimal point
\usepackage{bm}% bold math

\usepackage{hyperref}
\hypersetup{
 linktocpage = true,
     colorlinks,
    citecolor=blue,
    filecolor=black,
    linkcolor=blue,
    urlcolor=blue
}

\usepackage{color}
\usepackage{array}
\usepackage{amsmath}
\usepackage{physics}
\usepackage{xfrac}

\definecolor{darkgreen}{rgb}{0,0.5,0}

\newcommand{\cA}{\mathcal{A}}
\newcommand{\bZ}{\mathbb{Z}}

\usepackage{color}
\definecolor{royalblue}{rgb}{0.25,0.41,1}

\usepackage{pifont}

\begin{document}

\title{Monopole Catalysis from Standard Model Anomalies}

\author{Nathaniel Craig}
\affiliation{
 Kavli Institute for Theoretical Physics, Santa Barbara, CA 93106, USA
 }
\affiliation{Department of Physics, University of California, Santa Barbara, CA 93106, USA}

\author{Dan Sehayek}
\affiliation{Department of Physics, University of California, Santa Barbara, CA 93106, USA}

\date{\today}

\begin{abstract}
We study the generalized symmetries associated with Standard Model flavor and their fate in the presence of magnetic monopoles. This illuminates a striking phenomenological consequence: even when baryon and lepton number are classically conserved, the electroweak anomaly induces $\Delta B = \Delta L \in 3 \mathbb{Z}$ in monopole-fermion scattering for any monopole activating the anomaly. We demonstrate this mechanism in Pati-Salam theory and trinification. 
\end{abstract}

\maketitle

\section{Introduction}

The Standard Model's generalized global symmetries \cite{Gaiotto:2014kfa} imply novel phenomena associated with its possible UV completions. Nowhere is this more apparent than in the flavor sector, where the Standard Model enjoys a large classical 0-form global symmetry when the Yukawa couplings are taken to zero. It has long been appreciated that Adler--Bell--Jackiw (ABJ) anomalies with $SU(3)_C$ and $SU(2)_L$ break parts of this symmetry to discrete subgroups, while it was recently observed \cite{Cordova:2022qtz} that ABJ and 't Hooft anomalies also imbue this symmetry with higher-group \cite{Cordova:2018cvg} and non-invertible \cite{Choi:2022jqy, Cordova2023} structure that intertwines it with the magnetic 1-form symmetry of $U(1)_Y$. Such generalized symmetries carry emergence constraints that imply irreducible sources of flavor violation in UV completions of the Standard Model that break the magnetic 1-form symmetry. These constraints were articulated and interpreted classically in \cite{Cordova:2022qtz}, organizing the ways in which Standard Model fermions can be assembled into irreps of grand unified theories. In this paper we interpret the constraints quantum mechanically, organizing the violation of flavor symmetries in monopole-fermion scattering. 

This is particularly salient for the classical flavor symmetries of the Standard Model that survive when Yukawa couplings are turned on. Baryon number $B$ and lepton number $L$ are accidental classical 0-form global symmetries of the Standard Model whose combination $B+L$ is broken quantum mechanically by the electroweak anomaly to a $\bZ_{2 N_g}$ discrete symmetry \cite{Koren:2022bam, Wang:2022eag}. Coupled with $B-L$ conservation, this admits $B$- and $L$-violating processes satisfying the selection rule $\Delta B = \Delta L \in N_g \mathbb{Z}$, where $N_g = 3$ is the number of Standard Model generations. Weak instantons \cite{tHooft:1976rip, tHooft:1976snw} and sphalerons \cite{Klinkhamer:1984di,Kuzmin:1985mm,Rubakov:1996vz} have long been known to induce such $B+L$ violation, albeit with negligible rate at zero temperature. 

Far less appreciated is the possibility that monopoles participating in the electroweak anomaly mediate $\Delta B = \Delta L \in N_g \mathbb{Z}$ in monopole-fermion scattering.\footnote{This is in contrast to the well-known $\Delta B = \Delta L = 1$ Callan-Rubakov effect \cite{Rubakov:1981rg,Callan1982} in $SU(5)$ grand unified theories, where $U(1)_{B+L}$ is already broken to $\bZ_2$ classically by the exchange of $X,Y$ gauge bosons.} The possibility of $B+L$ violation in monopole-fermion scattering due to the {\it weak} anomaly was recognized long ago by Wilczek \cite{Wilczek:1981dr} and developed by Schellekens \cite{Schellekens:1984jf}, but remains woefully under-appreciated. The most persuasive demonstration of the effect is due to Sen \cite{Sen-1985-II,Sen-1985}, who showed that the $q_{\rm em}= 2$ magnetic monopole in Pati-Salam unification (which preserves a classical $U(1)_{B+L}$ symmetry acting on both matter fields and gauge bosons) supports fermion condensates with $\Delta B = \Delta L = 3$. Moreover, he demonstrated --- contrary to prior expectations \cite{Wilczek:1981dr} --- that this effect is unsuppressed by monopole size, coupling constants, or the breaking of electroweak symmetry. Beyond these early studies, however, the trail largely grows cold.\footnote{See \cite{Bennett:1985} for an early assessment of phenomenological limitations, and \cite{Brihaye:2002} for a later monopole--Skyrmion model of anomaly-induced baryon violation.} The generality of the mechanism and its potentially spectacular $\Delta B = \Delta L = 3$ signatures \cite{DiLuzio:2020xgc} have received remarkably little attention. 

As we will see, however, the modern language of generalized symmetries makes apparent that this phenomenon is not unique to Pati-Salam monopoles or the weak anomaly, but is instead generic: with rare exception, any monopole whose short-range magnetic fields participate in the weak or hypercharge ABJ anomalies of $U(1)_{B+L}$ should catalyze $\Delta B = \Delta L \in N_g \mathbb{Z}$ in monopole-fermion scattering. Since {\it every} monopole with a $U(1)_{\rm em}$ magnetic field at low energies supports a $U(1)_Y$ or $SU(2)_L$ magnetic field at high energies, $\Delta B = \Delta L \in N_g \mathbb{Z}$ in monopole-fermion scattering is an entirely ubiquitous property of magnetic monopoles.\footnote{The sole exception being monopoles supporting only the precise linear combination of $SU(2)_L$ and $U(1)_Y$ magnetic fields corresponding to $U(1)_{em}$, which does not participate in the anomaly. Even then, the necessary couplings can be furnished by the incident fermions.}

This paper is organized as follows: in Sections \ref{sec:2group} and \ref{sec:noninvt} we highlight the 2-group and non-invertible flavor symmetries of the Standard Model (following \cite{Cordova:2022qtz}) and interpret their attendant emergence constraints in terms of flavor violation in monopole-fermion scattering. In Section \ref{sec:sm-monopoles} we systematically study the effects of Standard Model ABJ anomalies in monopole backgrounds, relating recent developments leveraging generalized global symmetries to historical treatments. We then specialize in Section \ref{sec:guts} to the violation of $B+L$ in monopole-fermion scattering, demonstrating its genericity and exploring concrete realizations in classically $B+L$--preserving grand unified theories. This includes (to the best of our knowledge) the first identification of anomaly-induced monopole catalysis in trinification.

\section{2-group Symmetries and Emergence in the Standard Model} \label{sec:2group}

Our primary interest in this paper is the rich anomaly structure of the Standard Model's classical global $U(N_g)^5/U(1)_Y$ 0-form flavor symmetry, where $N_g = 3$ is the number of generations. For our purposes, it is convenient to separately discuss the $SU(N_g)_f$ and $U(1)_f$ factors of each $U(N_g)_f \simeq [SU(N_g)_f \times U(1)_f]/\mathbb{Z}_{N_g}$, recalling the common center where necessary. 

We begin with the mixed flavor-flavor-hypercharge anomaly coefficients (which, in a mild abuse of nomenclature, we will often refer to as 't Hooft anomalies; we will not be concerned with the genuine cubic flavor 't Hooft anomalies). The 't Hooft anomaly coefficients of the $SU(N_g)_f$ and $U(1)_f$ flavor symmetries with $U(1)_Y$ are enumerated in Table \ref{tab:thooft}. The fact that the $SU(N_g)_f^2 \times U(1)_Y$ anomaly coefficients are nonzero for all of $f\in\{Q,u,d,\ell,e\}$ implies that each of these non-abelian flavor symmetries participates in a 2-group with the hypercharge magnetic 1-form symmetry \cite{Cordova:2022qtz}. Were it not for additional ABJ anomalies, the same would be true of the $U(1)_f$ flavor symmetries, as we will discuss momentarily. 

\begin{table}[h!]\centering
\setlength{\extrarowheight}{2pt}
\large
\begin{tabular}{|c||c|c|} 
\hline
$f$ & $\cA_{SU(N_g)_f^2 U(1)_Y}$ & $\cA_{U(1)_f^2 U(1)_Y}$ \\[3pt] \hline \hline
$Q$ & $\sfrac{1}{2}$ & $N_g$ \\ \hline
$u_R$ & $-1$ & $ -2 N_g$ \\  \hline
$d_R$ & $\sfrac{1}{2}$ & $N_g$ \\ \hline
$\ell$ & $-\sfrac{1}{2}$ & $-N_g$ \\ \hline
$e_R$ & $\sfrac{1}{2}$ & $N_g$  \\ \hline
\end{tabular}
\caption{'t Hooft anomaly coefficients of the Standard Model flavor symmetries with $U(1)_Y$.}
\label{tab:thooft}
\end{table}

These 2-group structures have dynamical implications, particularly when both the gauge and global symmetries are emergent in the infrared \cite{Cordova:2018cvg}, as is often the case in extensions of the Standard Model. If $SU(N_g)_f$ emerges below a scale $E_f^{(0)}$ and the magnetic one-form symmetry of $U(1)_Y$ emerges below a scale $E_\text{mag}^{(1)}$, then the 2-group structure implies the emergence constraint
\begin{equation}
E_f^{(0)}\lesssim E_\text{mag}^{(1)} .
\end{equation}
Thus the non-Abelian flavor symmetry cannot emerge at a parametrically higher scale than the magnetic one-form symmetry. In \cite{Cordova:2022qtz}, this constraint was used to identify possible embeddings of Standard Model fermions into unified multiplets: in unified theories whose breaking to the Standard Model supports 't Hooft-Polyakov monopoles, the emergence constraint implies that only combinations of flavor symmetries with vanishing mixed hypercharge 't Hooft anomaly can persist above the unification scale. The violation of anomalous flavor symmetries was not tied directly to the monopoles breaking the magnetic 1-form symmetry, but rather classical symmetry breaking due to the exchange of massive gauge bosons at the unification scale. This exemplifies a common feature of weakly-coupled models, where phenomena satisfying the emergence constraint often appear well before the constraint is saturated \cite{Brennan:2020ehu, Choi:2022fgx, Sehayek:2026pvu}.

But the emergence constraint must be satisfied in any ultraviolet completion, which suggests phenomena that directly link violation of the one-form symmetry to violation of the zero-form symmetry. For example, the emergence constraint associated with the 3-group symmetry of axion EFTs is automatically satisfied by fermion zero modes living on axion strings \cite{Sehayek:2026pvu}. As for our 2-group symmetries, dynamical monopoles carrying hypercharge magnetic fields break the continuous magnetic one-form symmetry, so every $SU(N_g)_f$ that participates nontrivially in the two-group must also be broken at or below the monopole scale \cite{Cordova:2018cvg,Cordova:2022qtz}. Global symmetry violation in monopole-fermion scattering is the natural candidate, and indeed, in the radial theory the mixed 't Hooft anomaly becomes a boundary anomaly that forbids an $SU(N_g)_f$-preserving boundary condition at the monopole core \cite{Jensen2018,Thorngren2021}. Hypercharge monopoles therefore violate the corresponding non-Abelian flavor symmetries in monopole-fermion scattering \cite{vanBeest:2023dbu}, satisfying the emergence constraint implied by the 2-group structure. 

Thus we arrive at the general phenomenological expectation that $SU(N_g)_f$ flavor symmetries of the Standard Model are violated in scattering off hypercharge magnetic monopoles. As for the $U(1)_f$ symmetries, their fate depends additionally on their ABJ anomalies.

\section{Non-invertible symmetries and emergence in the Standard Model} \label{sec:noninvt}

In addition to their 't Hooft anomalies with $U(1)_Y$, the $U(1)_f$ flavor symmetries have ABJ anomalies with various factors of the Standard Model gauge group, as indicated in Table \ref{tab:abj}. 

\begin{table}[h!]\centering
\setlength{\extrarowheight}{2pt}
\large
\begin{tabular}{|c||c|c|c|}  \hline
$f$ & $\cA_{U(1)_f U(1)_Y^2}$ & $\cA_{U(1)_f SU(2)_L^2}$ & $\cA_{U(1)_f SU(3)_C^2}$ \\[2pt] \hline \hline
$Q$ & $\sfrac{N_g}{6}$ & $\sfrac{3 N_g}{2}$ & $N_g$ \\ \hline
$u_R$ & $-\sfrac{4 N_g}{3}$ & 0 & $-\sfrac{N_g}{2}$ \\ \hline
$d_R$ & $ - \sfrac{N_g}{3}$ & 0 & $-\sfrac{N_g}{2}$ \\ \hline
$\ell$ & $\sfrac{N_g}{2}$ & $\sfrac{N_g}{2}$ & 0 \\ \hline
$e_R$ & $ -N_g$ & 0 & 0 \\ \hline
\end{tabular}
\caption{ABJ anomaly coefficients of the Standard Model flavor symmetries with $SU(3)_C \times SU(2)_L \times U(1)_Y.$}
\label{tab:abj}
\end{table}

At the level of the Lie algebra, it is convenient to form linear combinations to diagonalize the ABJ anomalies. Two linear combinations are entirely anomaly-free, namely those corresponding to $U(1)_{B-L}$ and (gauged) hypercharge. As for the rest, the linear combinations $u_R - 4 d_R$, $2 \ell + e_R$, and $e_R$ carry ABJ anomalies with respect to $SU(3)_C$, $SU(2)_L$, and $U(1)_Y$, respectively. Of these, the surviving invertible subgroups depend on the global structure of the Standard Model gauge group $[SU(3)_c \times SU(2)_L \times U(1)_Y]/\Gamma$. The linear combination $U(1)_{u_R - 4 d_R}$ is broken to $\mathbb{Z}_{N_g}$ ($\mathbb{Z}_{3 N_g}$) for $\Gamma = \mathbb{Z}_3, \mathbb{Z}_6$ ($\mathbf{1}, \mathbb{Z}_2$), while the linear combination $U(1)_{2 \ell + e_R}$ is broken to $\mathbb{Z}_{N_g}$ ($\mathbb{Z}_{2 N_g}$) for $\Gamma = \mathbb{Z}_2, \mathbb{Z}_6$ ($\mathbf{1}, \mathbb{Z}_3$). In either remnant, the subgroup generated by a rotation through
$2\pi/N_g$ is already contained in the non-Abelian flavor centers:
it acts as $(\omega,\omega^{-4})$ in
$SU(N_g)_u\times SU(N_g)_d$ for $U(1)_{u_R-4d_R}$, and as
$(\omega^2,\omega)$ in $SU(N_g)_\ell\times SU(N_g)_e$ for
$U(1)_{2\ell+e_R}$. Thus the larger $\mathbb Z_{3N_g}$ and
$\mathbb Z_{2N_g}$ remnants contain, respectively, additional
$\mathbb Z_3$ and $\mathbb Z_2$ components beyond the flavor centers, whereas for $\Gamma=\mathbb Z_6$ no additional Abelian discrete factor survives.

As for $U(1)_{e_R}$, using the fractional-bundle quantization conditions of \cite{Cordova2024}, one finds $U(1)_e \rightarrow \{\bZ_{N_g}, \bZ_{4 N_g}, \bZ_{9 N_g}, \bZ_{36 N_g} \}$ for $\Gamma = \{\bZ_6, \bZ_3, \bZ_2, \mathbf{1} \}$. The rest of $U(1)_e$ is non-invertible \cite{Choi:2022jqy, Cordova2023}, making this the one proper non-invertible flavor symmetry of the Standard Model in the absence of Yukawa couplings. 

Of course, the story changes when Yukawa couplings are accounted for. When nonzero, these classically preserve the linear combination of $U(1)_f$ symmetries identified with $U(1)_B$ and, in the absence of Majorana neutrino masses, $U(1)_L$. As noted above, the linear combination $U(1)_{B-L}$ has no mixed anomalies with the Standard Model gauge group, while the orthogonal combination $U(1)_{B+L}$ has ABJ anomalies with both $SU(2)_L$ and $U(1)_Y$ that break $U(1)_{B+L} \rightarrow \mathbb{Z}_{2 N_g}$ regardless of the global structure of the gauge group. Together with $B-L$ conservation, the surviving discrete symmetry imposes the selection rule \cite{Koren:2022bam}
\begin{equation}
\Delta B=\Delta L\in N_g\mathbb Z.
\end{equation}
The $SU(2)_L$ anomaly removes the non-invertible symmetry of $U(1)_{B+L}$ that the hypercharge anomaly would otherwise support.
That being said, it is useful to consider the Standard Model in the formal limit of strong and weak couplings $g_s,g\rightarrow0$ with hypercharge coupling $g'$ fixed, for which the non-Abelian instanton effects vanish and $U(1)_{B+L}$ supports the non-invertible symmetry associated with its hypercharge ABJ anomaly. The value of this thought experiment is that it invites us to consider the implications of the corresponding emergence constraint.

What do these ABJ anomalies imply about the fate of $U(1)_{B+L}$ in monopole-fermion scattering? A monopole carrying magnetic charge under a Standard Model gauge factor activates the corresponding ABJ anomaly in the radial scattering problem and, as has long been appreciated, weak and color magnetic fields can therefore induce violation of flavor symmetries with $SU(2)_L$ and $SU(3)_C$ ABJ anomalies. In particular, a weak monopole activates the $B+L$ anomaly, and for $N_g=3$ the minimal nonzero process allowed by the above selection rule has
\begin{equation}
|\Delta B|=|\Delta L|=3.
\end{equation}

As for the effect of abelian ABJ anomalies on monopole-fermion scattering, here the value of our $g \rightarrow 0$ thought experiment is that it organizes the symmetry violation associated with scattering off hypercharge monopoles. The fusion rules for the non-invertible symmetry defects imply the same emergence constraints as that of the 2-group: $E_f^{(0)}\lesssim E_\text{mag}^{(1)}$. Consequently, monopoles carrying hypercharge magnetic fields violate flavor symmetries with $U(1)_Y$ ABJ anomalies in monopole-fermion scattering, including $U(1)_{B+L}$.

Since any monopole carrying electromagnetic flux below the scale of EWSB necessarily carries hypercharge flux above that scale, this implies that anomalous $B+L$ violation is a generic feature of monopole-fermion scattering in the Standard Model. For $N_g=3$, the minimal nonzero fermionic channel allowed by the anomaly has $|\Delta B|=|\Delta L|=3$ \cite{Bennett:1985}. This differs from the familiar Callan--Rubakov effect in $SU(5)$, where GUT gauge interactions classically violate $B+L$ and allow $|\Delta B|=|\Delta L|=1$. Here the selection rule follows from the electroweak anomaly together
with $B-L$ conservation, and therefore persists in unified theories
whose complete classical interactions preserve both $B+L$ and $B-L$,
provided their ultraviolet anomalies preserve the same discrete subgroup
of $B+L$.

For neutral Cartan backgrounds after electroweak symmetry breaking (EWSB), the $B+L$ anomaly becomes
\begin{equation}
\begin{aligned}
d \star J_{B+L} &= \frac{N_g}{8 \pi^2} \left[ g^2 F_{W^3}\wedge F_{W^3}
-g'^2F_Y\wedge F_Y \right] \\
&= \frac{N_g}{8 \pi^2} \left[2gg'F_\gamma\wedge F_Z
 +(g^2-g'^2)F_Z\wedge F_Z \right].
\end{aligned}
\end{equation}
The $F_\gamma\wedge F_\gamma$ term cancels, as required by the absence of a $B+L$ anomaly with electromagnetism. Nevertheless, the mixed term can be activated in scattering in either of two ways. A
monopole whose short-distance magnetic charge is not aligned with $t_{\rm em}$ carries a short-range $Z$ magnetic field, which pairs with the electromagnetic electric field of the incident fermion. Alternatively, the monopole's electromagnetic magnetic field can
pair with a $Z$ electric field sourced by the incident fermion or by a dyonic monopole.

While naively one might expect the nonzero $W, Z$ masses from EWSB to suppress the rate of monopole-fermion scattering from $\sim 1/\Lambda_{\rm QCD}^2$ to $\sim 1/m_W^2$ (or worse), Sen showed that when the monopole core radius satisfies
\begin{equation} \label{eq:unsuppressed}
r_0<\frac{g^2E}{m_W^2},
\end{equation}
where $E$ is the energy of the external fermions, the weak-anomaly contribution is not power suppressed by the monopole size, the electroweak scale, or the weak coupling \cite{Sen-1985-II}. This suggests the rate of $B+L$ violation through the electroweak anomaly is geometric, much as the familiar $SU(5)$ Callan-Rubakov effect. However, for $N_g=3$, the minimal anomalous operator contains fermions from all three generations. Depending on the scattering energy and available phase space, this may introduce suppression by powers of small mixing angles that depend on the details of the misalignment between the gauge and mass eigenstates of the Standard Model \cite{Bennett:1985}. Moreover, processes with $|\Delta B|=|\Delta L|=3$ require a large number of Standard Model fermions, implying significant phase space suppression unless the initial state carries appreciable baryon number. 

These considerations lead us to expect that $U(1)_{B+L}$ violation in monopole-fermion scattering with $|\Delta B|=|\Delta L|=3$ is a generic effect of monopoles carrying microscopic $SU(2)_L$ or $U(1)_Y$ magnetic fields, a potentially striking phenomenological signature that has not been extensively recognized or studied \footnote{We reserve a more detailed study of $|\Delta B|=|\Delta L|=3$ phenomenology in monopole-fermion scattering for forthcoming work \cite{CraigGao}.}.  While this expectation follows on very general grounds from the anomaly structure of the Standard Model 0-form flavor symmetries, we can demonstrate this phenomenon more concretely by studying magnetic monopoles in grand unified theories where $U(1)_{B+L}$ is a valid classical symmetry: Pati-Salam and trinification. Before we do so, however, it is useful to make contact between our expectations for monopole-fermion scattering from ABJ anomalies in the unbroken phase of the Standard Model and the historical treatment of anomalous monopole-fermion scattering in monopole backgrounds.

\section{Anomalies in Standard Model Monopole Backgrounds} \label{sec:sm-monopoles}

\subsection{Anomalies in the Monopole Background} 

At large radius, the magnetic field of an isolated monopole is characterized by a $U(1)_M$ embedding in the unbroken gauge group $G$, generated by a Dirac-quantized magnetic charge $Q_{\rm mg}$ that may be chosen in the Cartan subalgebra of $G$. In a fixed monopole background, the gauge transformations acting conventionally on scattering states are those commuting with $Q_{\rm mg},$ namely its centralizer. Fermions decompose into weights $w$ of $G$, and each component sees an ordinary abelian magnetic charge $q_M = w(Q_{\rm mg}).$ This makes it natural to discuss monopole-fermion scattering in terms of the unbroken Cartan charges in a given monopole embedding. The anomalies involving \(U(1)_M\) and the Cartan gauge factors preserved by a monopole are the monopole-background reductions of the gauge-invariant bulk anomalies. The bulk anomaly identifies the possible obstruction, while contraction with \(Q_{\rm mg}\) determines whether a specified monopole activates it. Upon reduction to the radial \(1+1\)-dimensional theory, this obstruction becomes a ’t Hooft anomaly that prevents an ordinary symmetry-preserving boundary condition at the monopole core. 

For a component of a Weyl fermion with definite monopole charge $q_{M,f}\neq0$, the Dirac operator on the sphere surrounding the monopole has $|q_{M,f}|$ zero modes. These form a single angular-momentum multiplet with
\begin{equation}
2j_0+1=|q_{M,f}|,
\end{equation}
and reduce to $|q_{M,f}|$ chiral fermion channels in the radial theory with chirality $(-1)^{\chi_f}\operatorname{sgn}(q_{M,f})$ \cite{vanBeest:2023dbu}, where $\chi_f=0$ for $f\in\{Q,\ell\}$ and $\chi_f=1$ for $f\in\{u_R,d_R,e_R\}$. Consequently, the two-dimensional anomaly coefficient for Abelian charges $X$ and $Y$ is
\begin{equation}
\mathcal A_{XY}^{(2d)}
=\sum_f(-1)^{\chi_f}q_{M,f}X_fY_f.
\end{equation}
This is precisely the contraction of the four-dimensional triangle anomaly with $Q_{\rm mg}$.

In the case of the Standard Model above the scale of EWSB, the monopole charge matrix (generator of $U(1)_M$) for stable monopoles must generally be of the form 
\begin{equation}
    Q_\text{mg}=\alpha t_Y+\beta t_L^3+\gamma t_C^8
\end{equation} 
where $t_Y$ is the hypercharge generator, $t_L^3=\text{diag}(1/2,-1/2)\in\mathfrak{su}(2)_L$ is the $U(1)_I$ (weak isospin) generator, and $t_C^8=\text{diag}(1/3,1/3,-2/3)\in\mathfrak{su}(3)_C$ is the $U(1)_{Y_C}$ (color hypercharge) generator. Charged gauge-field fluctuations destabilize a monopole whenever a
$W$ boson or gluon sees an effective magnetic charge $|q_M|\geq2$;
the resulting vector-boson condensation screens the nonminimal
non-Abelian magnetic charge \cite{Brandt:1979kk,vanBeest:2023mbs}.
After using Weyl transformations, the stable representatives therefore
satisfy the Brandt--Neri stability criteria
\begin{equation}
\beta\in\{0,1\},
\qquad
\gamma\in\{0,\pm1\}.
\end{equation}
 These coefficients are further related by Dirac quantization, which requires
\begin{equation}
\beta\equiv\alpha\pmod 2,
\qquad
\gamma\equiv\alpha\pmod 3.
\end{equation}
Finally, the allowed values of $\alpha$ depend on the global form of the Standard Model gauge group.

We can now ask whether a given monopole violates flavor symmetries in monopole-fermion scattering by relating the Cartan anomalies associated with its magnetic field to the bulk ABJ anomalies. Let's begin with the $U(1)_f\times U(1)_M\times U(1)_Y$ anomaly, whose coefficient is
\begin{align}
    \mathcal{A}_f^{(Y)}
    &=(-1)^{\chi_f}\sum_{i\in f}
      \left(\alpha q_{i,Y}+\beta q_{i,I}+\gamma q_{i,Y_C}\right)q_{i,Y}\\
    &=(-1)^{\chi_f}\sum_{i\in f}\alpha q_{i,Y}^2
      =\alpha\cA_{U(1)_f U(1)_Y^2},
\end{align}
where $q_{i,X}$ is the $U(1)_X$ charge of particle $i$. This follows since the color hypercharges cancel for $SU(3)_C$ triplets, weak isospin charges cancel for $SU(2)_L$ doublets, and are zero otherwise. By linearity, the same relation holds for any linear combination of the five $U(1)_f$ factors. Thus a monopole with $\alpha\neq0$ activates precisely the radial anomalies inherited from the bulk $U(1)_f U(1)_Y^2$ ABJ anomalies, forcing violation of the corresponding Abelian flavor charges in monopole-fermion scattering. This is the monopole-scattering mechanism anticipated by the emergence constraint $E_\text{flavor}^{(f)}\lesssim E_\text{mag}^{(Y)}$ discussed above.

An analogous statement holds for the $U(1)_f\times U(1)_M\times U(1)_I$ anomalies. For $f\in\{Q,\ell\}$, the corresponding anomaly coefficient is
\begin{align}
    \mathcal{A}_f^{(I)}
    &=\sum_{i\in f}
      \left(\alpha q_{i,Y}+\beta q_{i,I}+\gamma q_{i,Y_C}\right)q_{i,I}\\
    &=\sum_{i\in f}\beta q_{i,I}^2
      =\beta\cA_{U(1)_f SU(2)_L^2}.
\end{align}
since hypercharge is constant across each doublet, while the color and weak traces factorize and are separately traceless. Consequently, a monopole with $\beta \neq 0$ activates the same anomaly carried by the bulk $U(1)_f SU(2)_L^2$ triangle. This mechanism can violate $U(1)_Q, U(1)_\ell$, and anomalous linear combinations thereof in monopole-fermion scattering.  

Likewise, for $f\in\{Q,u_R,d_R\}$, the $U(1)_f\times U(1)_M\times U(1)_{Y_C}$ anomaly coefficient is
\begin{align}
\mathcal A_f^{(Y_C)}
&=(-1)^{\chi_f}\sum_{i\in f}
  \left(\alpha q_{i,Y}+\beta q_{i,I}
  +\gamma q_{i,Y_C}\right)q_{i,Y_C}\\
&=(-1)^{\chi_f}\gamma\sum_{i\in f}q_{i,Y_C}^2
  =\frac{4\gamma}{3}\cA_{U(1)_fSU(3)_C^2}.
\end{align}
Thus a monopole with $\gamma \neq 0$ activates the radial reduction of the corresponding bulk $U(1)_fSU(3)_C^2$ ABJ anomaly.

\subsection{Monopoles in and beyond the Standard Model}

Having translated the necessary conditions for anomalous violation of Standard Model flavor symmetries by monopole-fermion scattering into the monopole charge matrix, it's now useful to ask what charge matrices we expect in the Standard Model both before and after EWSB.

The global form of the Standard Model gauge group before electroweak symmetry breaking is
\begin{equation}
    G_{\rm SM}^{(\Gamma)}=
    \frac{SU(3)_C\times SU(2)_L\times U(1)_{\widehat Y}}{\Gamma},
    \qquad \widehat Y=6Y,
\end{equation}
where $\Gamma\in\{1,\mathbb Z_2,\mathbb Z_3,\mathbb Z_6\}$.  Given a global structure, the Dirac condition is that \(\exp(2\pi iQ_{\rm mg})\) be the identity in \(G_{\rm SM}^{(\Gamma)}\), or equivalently an element of \(\Gamma\) when lifted to the covering product.  Each allowed magnetic one-form charge has a Brandt--Neri-stable representative \cite{Brandt:1979kk,vanBeest:2023mbs} with
\begin{equation}
    \alpha\in \frac{6}{|\Gamma|}\mathbb Z.
\end{equation}
The first six positive representatives are displayed in Table~\ref{tab:monopoles-pre-EWSB}.  The pattern repeats with period six in the non-Abelian components.

\begin{table}[!htbp]
\centering
\begin{tabular}{|l||c|c|c|c|}
\hline
\(Q_{\rm mg}\) & \(\Gamma=1\) & \(\Gamma=\mathbb Z_2\) & \(\Gamma=\mathbb Z_3\) & \(\Gamma=\mathbb Z_6\) \\ \hline \hline
\(t_Y+t_L^3+t_C^8\)       & -- & -- & -- & \checkmark \\ \hline
\(2t_Y-t_C^8\)             & -- & -- & \checkmark & \checkmark \\ \hline
\(3t_Y+t_L^3\)             & -- & \checkmark & -- & \checkmark \\ \hline
\(4t_Y+t_C^8\)             & -- & -- & \checkmark & \checkmark \\ \hline
\(5t_Y+t_L^3-t_C^8\)       & -- & -- & -- & \checkmark \\ \hline
\(6t_Y\)                    & \checkmark & \checkmark & \checkmark & \checkmark \\ \hline
\end{tabular}
\caption{Brandt--Neri-stable magnetic charges with
$\alpha=1,\ldots,6$. A checkmark indicates that the charge satisfies
Dirac quantization for the indicated global form of the Standard Model
gauge group.}
\label{tab:monopoles-pre-EWSB}
\end{table}

Note that Dirac quantization classifies allowed 't Hooft lines of the low-energy gauge group, while topological monopoles in a breaking \(G_{\rm UV}\to G_{\rm SM}^{(\Gamma)}\) are classified by
\begin{equation}
 \pi_2\!\left(G_{\rm UV}/G_{\rm SM}^{(\Gamma)}\right)
 \simeq
 \ker\!\left[\pi_1\!\left(G_{\rm SM}^{(\Gamma)}\right)
 \longrightarrow \pi_1(G_{\rm UV})\right]
 \label{eq:uv-monopole-kernel}
\end{equation}
when \(\pi_2(G_{\rm UV})=0\) \cite{Goddard:1978}.  Thus an allowed 't Hooft charge is realized as a topologically stable soliton only when its loop is contractible in the chosen ultraviolet gauge group.

After electroweak symmetry breaking, a deconfined magnetic line has no long-range $Z$ component, so its weak and hypercharge coefficients agree. Thus
\begin{equation}
Q_{\rm mg}=n t_{\rm em}+\gamma t_C^8,
\qquad
t_{\rm em}=t_Y+t_L^3,
\qquad
n=\alpha.
\end{equation}
The spectrum of genuine magnetic lines continues to depend on the pre-EWSB global form:
\begin{equation}
n\in\frac{6}{|\Gamma|}\mathbb Z,
\qquad
\gamma\equiv n\pmod 3.
\end{equation}
In particular, the $n=2$ and $n=3$ sectors have charges
$2t_{\rm em}-t_C^8$ and $3t_{\rm em}$, respectively
\cite{Tong:2017oea,Dawson:1982sc,vanBeest:2023mbs,
Khoze:2024hlb,Alonso:2025cq}.

These possibilities can be related to familiar ultraviolet
constructions \cite{Tong:2017oea,Cordova2024,Alonso:2025cq}.
The conventional Georgi--Glashow $SU(5)$ embedding gives
$\Gamma=\mathbb Z_6$, and its elementary smooth monopole has
$n=1$. Direct-product Pati--Salam,
\begin{equation}
SU(4)_C\times SU(2)_L\times SU(2)_R,
\end{equation}
gives $\Gamma=\mathbb Z_3$, and its elementary smooth monopole has
$n=2$. The quotient
\begin{equation}
\frac{SU(4)_C\times SU(2)_L\times SU(2)_R}{\mathbb Z_2},
\end{equation}
as inherited from a $Spin(10)$ embedding, gives
$\Gamma=\mathbb Z_6$. Nevertheless, Eq.~\eqref{eq:uv-monopole-kernel}
implies that Pati--Salam breaking produces only even-winding smooth
monopoles \cite{Goddard:1978}. The $n=1$ magnetic line
is allowed by the low-energy global form \cite{Alonso:2025cq}, but realizing it as a smooth monopole requires a core inherited from an earlier $Spin(10)$
breaking transition.

Direct-product trinification gives $\Gamma=\mathbb Z_2$, and its
elementary smooth monopole has $n=3$. The $E_6$-compatible
trinification group
\begin{equation}
\frac{SU(3)_C\times SU(3)_L\times SU(3)_R}{\mathbb Z_3}
\end{equation}
instead induces $\Gamma=\mathbb Z_6$, but trinification breaking
itself produces only $n\in3\mathbb Z$. The $n=1$ and $n=2$ sectors
require monopole cores inherited from an earlier $E_6$ breaking
transition.

Among the positive, Brandt--Neri-stable pre-EWSB charges, only
$n=1$ has its electroweak component aligned with $t_{\rm em}$,
so the elementary Pati--Salam and trinification monopoles carry
short-range $Z$ magnetic fields after EWSB.
This parallels Callan's observation that the elementary $SU(5)$
dyon carries no $Z$ electric charge whereas the Pati--Salam dyon
can, providing a source for anomaly-induced baryon number
violation even when baryon number is conserved classically
\cite{Callan1984Summary}.

\section{Anomalous $B+L$ violation in Grand Unified Theories} \label{sec:guts}

We now turn to classically $B+L$-conserving grand unified theories whose monopoles are expected, by our general arguments, to catalyze $|\Delta B| = |\Delta L| = 3$ in monopole-fermion scattering. This has the advantage of allowing us to construct, to varying degrees, concrete examples demonstrating the effect. It also significantly broadens the scope of monopole-induced $B+L$ violation in grand unified theories beyond the familiar $|\Delta B| = |\Delta L| = 1$ Callan-Rubakov effect in $SU(5)$.

We start with Pati--Salam, analyzing the $q_Y=2$ Pati--Salam monopole and its $q_{\rm em}=2$ continuation after EWSB.  Following Sen \cite{Sen-1985}, we identify monopole-induced fermion condensates by embedding the magnetic charge in an $SU(2)_M$ and constructing the non-anomalous phase rotations of the resulting fermion multiplets.  Because Pati--Salam breaking is generated by a non-adjoint Higgs, we verify that its vacuum expectation value lies in the adjoint-plus-singlet sector of the $SU(2)_M$ branching, as required by the construction of Ref.~\cite{Laterveer:1988}. We then turn to trinification, where hypercharge monopoles admit no spherically symmetric embedding and the condensate cannot be determined by the same partial-wave analysis, but analogous expectations for monopole catalysis continue to hold.

\subsection{Pati-Salam} \label{sec:PS}

Let us first consider the $q_Y=2$ monopole of Pati-Salam, which has charge matrix $Q_\text{mg}=2t_Y-t_C^8$. Unlike the case of $SU(5)$, Higgsing of the Pati-Salam gauge group to $G_\text{SM}$ is accomplished using a \textit{non-adjoint} Higgs. In particular, we may take the Higgs field to live in either the $(\overline{\mathbf{10}},\mathbf{1},\mathbf{3})$ or the $(\overline{\mathbf{4}},\mathbf{1},\mathbf{2})$ of $SU(4)_C\times SU(2)_L\times SU(2)_R$. Taking
\begin{equation}
t_{B-L}
=
\operatorname{diag}\left(
\frac{1}{3},\frac{1}{3},\frac{1}{3},-1
\right),
\end{equation}
the branching of the rank-two symmetric representation
$\overline{\mathbf{10}}$ under
$SU(3)_C\times U(1)_{B-L}$ is
\begin{equation}
\overline{\mathbf{10}}
\rightarrow
\overline{\mathbf6}(-2/3)
\oplus\overline{\mathbf3}(+2/3)
\oplus\mathbf1(+2).
\end{equation}
The corresponding branching of the antifundamental
$\overline{\mathbf4}$ is
\begin{equation}
\overline{\mathbf4}
\rightarrow
\overline{\mathbf3}(-1/3)
\oplus
\mathbf1(+1).
\end{equation}
In either case, a VEV for the Standard Model singlet breaks
$SU(4)_C\times SU(2)_R$ to $SU(3)_C\times U(1)_Y$, where
\begin{equation}
t_Y=t_R^3+\frac12t_{B-L},
\qquad
t_R^3=\frac12\operatorname{diag}(1,-1).
\end{equation}

Unlike the conventional adjoint-Higgs 't Hooft--Polyakov construction, Pati--Salam breaking is driven by a non-adjoint Higgs,
so the usual Bogomol'nyi equation does not apply directly. As described in \cite{Laterveer:1988}, it is however possible to obtain a Bogomol’nyi equation in this more general case, presuming that the branching of the Higgs with respect to the chosen embedding includes an adjoint representation, and that it is only this representation (in addition to possible singlets) that acquires a vacuum expectation value. In the Bogomol'nyi limit, and provided the embedded fields form a consistent truncation, minimization of the energy gives the usual monopole equation for the adjoint component. For the case in which the Higgs field lives in the $(\overline{\mathbf{4}},\mathbf{1},\mathbf{2})$ of the Pati-Salam gauge group, we find that the following $SU(2)_M$ embedding produces the desired monopole charge matrix and satisfies this property:
\begin{equation}
    t_M^a=t_R^a+t_{(34)}^a \label{eq:qy2-embedding}
\end{equation}
where $t_R^a$ are the generators of $SU(2)_R$ and $t_{(34)}^a$ are the generators of the $SU(2)$ subgroup spanning the $3-4$ space of $SU(4)_C$. Under the diagonal $SU(2)_M$ embedding,
\begin{align}
(\overline{\mathbf4},\mathbf1,\mathbf2)
&\rightarrow
(\mathbf2\oplus2\,\mathbf1)\otimes\mathbf2\\
&=
\mathbf3\oplus\mathbf1\oplus2\,\mathbf2.
\end{align}
The Standard Model singlet that acquires the VEV lies in $\mathbf2\otimes\mathbf2$ and has components in the $\mathbf3\oplus\mathbf1$. The VEV therefore satisfies the adjoint-plus-singlet condition of Ref.~\cite{Laterveer:1988}. In the first-stage Bogomol'nyi limit with equal $SU(4)_C$ and $SU(2)_R$ gauge couplings, the diagonal ansatz consistently embeds a spherically symmetric 't Hooft--Polyakov monopole.

The same criterion fails for a Higgs in
$(\overline{\mathbf{10}},\mathbf1,\mathbf3)$. Its full branching is
\begin{align}
(\overline{\mathbf{10}},\mathbf1,\mathbf3)
&\rightarrow
(\mathbf3\oplus2\,\mathbf2\oplus3\,\mathbf1)\otimes\mathbf3\\
&=
\mathbf5\oplus2\,\mathbf4\oplus4\,\mathbf3
\oplus2\,\mathbf2\oplus\mathbf1.
\end{align}
The Standard Model singlet VEV has a nonzero projection onto the $\mathbf5$, so it does not satisfy the criterion of
\cite{Laterveer:1988}. We therefore use the $(\overline{\mathbf4},\mathbf1,\mathbf2)$ Higgs in the analysis below.

This symmetry-breaking sector can preserve both $U(1)_{B+L}$ and $U(1)_{B-L}$ at the classical level. Start with a $U(1)_{F}$ global symmetry that assigns unit charge to both matter fields $F_L\sim(\mathbf4,\mathbf2,\mathbf1)_L$ and
$F_R\sim(\mathbf4,\mathbf1,\mathbf2)_R$ and a $U(1)_{\sigma}$ global symmetry that assigns unit charge to the GUT-breaking Higgs $\Sigma\sim(\overline{\mathbf4},\mathbf1,\mathbf2)$. Taking the scalar potential to preserve $U(1)_{\sigma}$ and the electroweak bidoublet containing the Standard Model Higgs to be neutral under both $U(1)_{F}$ and $U(1)_\sigma$, define
\begin{equation}
B+L=\frac12\left(F+\sigma-t_{B-L}\right),
\qquad
B-L=t_{B-L}-\sigma.
\end{equation}
On the fermions these are the usual $B+L$ and $B-L$ charges. The Standard Model-singlet component of $\Sigma$ has $(\sigma,t_{B-L})=(1,1)$, so its VEV preserves both combinations, as do the Dirac Yukawa couplings.

To determine the radial condensate, we follow Sen's fixed-background, lowest-partial-wave treatment and neglect Yukawa couplings and gauge-field fluctuations that mix different $SU(2)_M$ multiplets. In this reduced radial theory, each charged $SU(2)_M$ multiplet has an independent phase rotation. With the 't Hooft--Polyakov embedding identified, we determine which of these phase rotations remain
conserved in the monopole background. Following \cite{Sen1984}, the mixed anomalies with every Cartan gauge generator commuting with $Q_{\rm mg}$ must be included. Let $U(1)_{j,a}$ rotate the $j$th $SU(2)_M$ multiplet in generation $a$. For a Cartan generator $g$, define
\begin{equation}
\mathcal A_{g,(j,a)}
=
\sum_{\psi\in(j,a)}
(-1)^{\chi_\psi}q_{M,\psi}q_{g,\psi}.
\end{equation}
A linear combination
\begin{equation}
S=\sum_{j,a}s_{j,a}U(1)_{j,a}
\end{equation}
is conserved precisely when
\begin{equation}
\sum_{j,a}\mathcal A_{g,(j,a)}s_{j,a}=0
\end{equation}
for every $g$. If there are $n_M$ multiplets per generation, then
$\mathcal A$ is a $4\times3n_M$ matrix in the examples below, and
\begin{equation}
\dim\ker(\mathcal A)
=
3n_M-\operatorname{rank}(\mathcal A).
\end{equation}
The gauge Cartans impose separate neutrality conditions and are not
included among these global conserved charges.

Let us now return to our $q_Y=2$ example. The branching rules of the fermion multiplets with respect to $SU(2)_M$ are
\begin{equation}
    (\mathbf{4},\mathbf{2},\mathbf{1})\rightarrow2\times\mathbf{2}\oplus4\times\mathbf{1}  ,\quad (\mathbf{4},\mathbf{1},\mathbf{2})\rightarrow\mathbf{3}\oplus\mathbf{2}\oplus\mathbf{2}\oplus\mathbf{1}
\end{equation}
The left-handed doublets are
\begin{equation}
\begin{pmatrix}u_{3L}\\ \nu_L\end{pmatrix},
\qquad
\begin{pmatrix}d_{3L}\\ e_L\end{pmatrix},
\label{eq:doublets-ps-qy2-I}
\end{equation}
while the right-handed multiplets are
\begin{equation}
\begin{pmatrix}u_{1R}\\ d_{1R}\end{pmatrix},
\qquad
\begin{pmatrix}u_{2R}\\ d_{2R}\end{pmatrix},
\qquad
\begin{pmatrix}
u_{3R}\\
2^{-1/2}(d_{3R}+\nu_R)\\
e_R
\end{pmatrix}.
\label{eq:doublets-ps-qy2-II}
\end{equation}
The first four doublets have magnetic charges $q_M=\pm1$. In each triplet, the $t_M^3=0$ component has $q_M=0$, while the $t_M^3=\pm1$ components have $q_M=\pm2$ and each contribute two lowest-partial-wave modes. We retain the $t_M^3=\pm1$ components as the fifth ``doublet.''

Let $N_{i,\alpha}^{(a)}$ denote the signed occupation number of component $\alpha$ in charged channel $i$, where $i=1,\ldots,4$ labels the four doublets, $i=5$ labels the $t_M^3=\pm1$ components of the triplet, and $a=1,2,3$ labels the generation. With this notation, one can verify that
\begin{equation}
    S_{i,j}^{(a)}=N_{i,1}^{(a)}+N_{i,2}^{(a)}-N_{j,1}^{(a)}-N_{j,2}^{(a)}
\label{eq:non-anomalous}
\end{equation}
for $(i,j)\in\{(1,2),(3,4)\}$ are non-anomalous.
For one generation, the anomaly matrix in the ordering of
Equations~(\ref{eq:doublets-ps-qy2-I}--\ref{eq:doublets-ps-qy2-II}) is
\begin{equation}
\mathcal A^{(a)}
=
\begin{pmatrix}
0 & 0 & 0 & 0 & 0\\
\frac{2}{3} & \frac{2}{3} & -1 & -1 & -\frac{10}{3}\\
-\frac{2}{3} & -\frac{2}{3} & 0 & 0 & \frac{4}{3}\\
0 & 0 & 0 & 0 & 0
\end{pmatrix},
\end{equation}
where the rows correspond to $I_L^3$, $Y$, $Y_C$, and $I_C^3$, and the relative sign of the right-handed fermions has been included. The full three-generation anomaly matrix therefore has rank two, so
\begin{equation}
\dim\ker(\mathcal A)=15-2=13.
\end{equation}

In addition to the six charges $S_{1,2}^{(a)}$ and $S_{3,4}^{(a)}$, there are six independent generation-difference
charges of the form
\begin{equation}
S_i^{(a)}=n_i^{(a)}-n_i^{(1)},
\qquad a=2,3,
\end{equation}
where \begin{equation}
n_i^{(a)}=N_{i,1}^{(a)}+N_{i,2}^{(a)}
\end{equation}
and we must account for the relations
\begin{equation}
S_i^{(a)}
=
S_j^{(a)}+S_{i,j}^{(a)}-S_{i,j}^{(1)},
\qquad
(i,j)\in\{(1,2),(3,4)\}.
\end{equation}
The remaining conserved charge may be chosen as
\begin{equation}
S_\star
=
\sum_{a=1}^3
\left(
n_1^{(a)}+n_2^{(a)}
-n_3^{(a)}-n_4^{(a)}
+n_5^{(a)}
\right).
\end{equation}
These thirteen charges form a basis for $\ker(\mathcal A)$; a fermion condensate must be neutral under all of them. A condensate invariant under the gauge symmetry preserved by the monopole and neutral under all thirteen conserved charges is
\begin{equation}
\mathcal O_{\rm PS}
=
\prod_{a=1}^3
\epsilon^{ij}
\bigl(u_{iR}^{(a)}u_{3R}^{(a)}\bigr)
\bigl(d_{jR}^{(a)}e_R^{(a)}\bigr),
\qquad i,j=1,2,
\label{eq:PS-condensate}
\end{equation}
where Lorentz and angular-momentum contractions are implicit. The operator is neutral
under the conserved phase rotations and all unbroken gauge generators,
while carrying
\begin{equation}
B=L=3.
\end{equation}

The conserved charges also determine the baryon and lepton numbers of
any monopole-induced condensate built from the charged channels above, and support our expectation that baryon and lepton number are each violated by (integer multiples of) three units in general.
In particular, they imply
\begin{equation}
n_1^{(a)}=n_2^{(a)}=A,
\qquad
n_3^{(a)}=n_4^{(a)}=N,
\qquad
n_5^{(a)}=C,
\end{equation}
independently of $a$. Neutrality under $S_\star$ then gives
\begin{equation}
C=2(N-A).
\end{equation}
Let $N_3$ denote the total signed occupation number of third-color
quarks in channels $1$, $2$, and $5$. Color-hypercharge neutrality
requires
\begin{equation}
0=G_{Y_C}
=\frac{1}{3}(6N)-\frac{2}{3}N_3
\quad\Longrightarrow\quad
N_3=3N.
\end{equation}
The total occupation number in channels $1$, $2$, and $5$ is therefore
\begin{equation}
\sum_{a=1}^3
\left(n_1^{(a)}+n_2^{(a)}+n_5^{(a)}\right)
=
3(2A+C)=6N.
\end{equation}
The remaining $6N-N_3=3N$ states in these channels are leptons.
Including the $6N$ color-1 and color-2 quarks in channels $3$ and $4$,
we obtain
\begin{equation}
\begin{aligned}
\Delta B&=\frac{1}{3}(6N+N_3)=3N,\\
\Delta L&=6N-N_3=3N,
\qquad N\in\mathbb Z.
\end{aligned}
\end{equation}
Thus the minimal nonzero condensate has
\begin{equation}
|\Delta B|=|\Delta L|=3,
\end{equation}
as realized by Eq.~(\ref{eq:PS-condensate}).

After EWSB, the monopole studied
above becomes the $q_{\rm em}=2$ monopole considered by Sen
\cite{Sen-1985}, with
\begin{equation}
Q_{\rm mg}=2t_{\rm em}-t_C^8=2t_M^3,
\qquad
t_M^a=t_L^a+t_R^a+t_{(34)}^a.
\end{equation}
The second stage of symmetry breaking $SU(3)_C \times SU(2)_L \times U(1)_Y \rightarrow SU(3)_C \times U(1)_{\rm em}$ is generated by a bifundamental
Higgs in $(\mathbf1,\mathbf2,\mathbf2^*)$, whose branching under
$SU(2)_M$ is
\begin{equation}
(\mathbf1,\mathbf2,\mathbf2^*)
\rightarrow\mathbf3\oplus\mathbf1.
\end{equation}

The fermion branchings are
\begin{equation}
(\mathbf4,\mathbf2,\mathbf1)_L
\rightarrow\mathbf3\oplus2\,\mathbf2\oplus\mathbf1,
\qquad
(\mathbf4,\mathbf1,\mathbf2)_R
\rightarrow\mathbf3\oplus2\,\mathbf2\oplus\mathbf1.
\end{equation}
The corresponding charged multiplets are
\begin{align}
F_L:\quad&
\begin{pmatrix}u_{1L}\\d_{1L}\end{pmatrix},\quad
\begin{pmatrix}u_{2L}\\d_{2L}\end{pmatrix},\quad
\begin{pmatrix}
u_{3L}\\
2^{-1/2}(d_{3L}+\nu_L)\\
e_L
\end{pmatrix},\\
F_R:\quad&
\begin{pmatrix}u_{1R}\\d_{1R}\end{pmatrix},\quad
\begin{pmatrix}u_{2R}\\d_{2R}\end{pmatrix},\quad
\begin{pmatrix}
u_{3R}\\
2^{-1/2}(d_{3R}+\nu_R)\\
e_R
\end{pmatrix}.
\end{align}
The four doublets have $q_M=\pm1$. In each triplet the middle
component has $q_M=0$, while the $t_M^3=\pm1$ components have $q_M=\pm2$ and
again supply two lowest-partial-wave modes.

Sen's conservation-law analysis gives
\begin{equation}
\Delta B=\Delta L\in3\mathbb Z,
\end{equation}
and he realizes the condensate in Eq.~(\ref{eq:PS-condensate}) as a minimal nonzero channel.

Restoring Yukawa couplings that preserve $B+L$ and $B-L$ breaks the additional flavor charges and rotates the condensate into
the mass basis, but leaves the selection rule $\Delta B=\Delta L\in N_g\mathbb Z$ unchanged. The rate may then depend on flavor mixing, as discussed above.

\subsection{Trinification} \label{sec:monopoles-trin}

We now turn to trinification, whose monopoles (to our knowledge) have not previously been identified with anomaly-induced $B+L$ violation. The elementary trinification monopole is necessarily
nonspherical, so unlike in Pati--Salam we cannot construct its fermion
condensate using Sen's partial-wave analysis. We can, however,
construct a theory of trinification that classically preserves both $B+L$ and $B-L$, show that this extension of Standard Model $B+L$ remains anomalous,
and verify that the elementary trinification monopole activates this anomaly with the expected selection rule. The lack of spherical symmetry does not appear to parametrically suppress the corresponding scattering process.

The two global forms of trinification relevant here are
\begin{equation}
\begin{aligned}
G_{\rm tri}^{\rm dir}
&=SU(3)_C\times SU(3)_L\times SU(3)_R,\\
G_{\rm tri}^{E_6}
&=\frac{SU(3)_C\times SU(3)_L\times SU(3)_R}{\mathbb Z_3},
\end{aligned}
\end{equation}
where the quotient is by the diagonal center. They induce the Standard
Model global forms $\Gamma=\mathbb Z_2$ and $\Gamma=\mathbb Z_6$,
respectively \cite{Alonso:2025cq}. Both may be broken to the Standard
Model by two Higgs fields
$\Phi_{1,2}\sim(\mathbf1,\mathbf3,\overline{\mathbf3})$, with
\begin{equation}
\langle\Phi_1\rangle=v_1E_{33},
\qquad
\langle\Phi_2\rangle=v_2E_{32},
\end{equation}
where $E_{ij}$ has a single nonzero entry \cite{Babu:1986}. The
unbroken hypercharge generator is
\begin{equation}
\begin{aligned}
t_{Y,L}&=\frac16\operatorname{diag}(-1,-1,2),\\
t_{Y,R}&=\frac13\operatorname{diag}(-2,1,1),
\end{aligned}
\end{equation}
acting on a bifundamental as
$t_Y\Phi=t_{Y,L}\Phi-\Phi t_{Y,R}$.

For either global form, the kernel in
Eq.~\eqref{eq:uv-monopole-kernel} is $3\mathbb Z$. The elementary
smooth monopole therefore has asymptotic charge
\begin{equation}
Q_{\rm mg}^{(3)}=3t_Y+t_L^3
\end{equation}
before EWSB and $Q_{\rm mg}^{(3)}=3t_{\rm em}$ afterward
\cite{Lazarides:1989,Alonso:2025cq}.\footnote{Baryon-number-violating scattering for this magnetic
charge was also considered in Ref.~\cite{Khoze:2024hlb}, using
a single fermion generation. In the classically $B$- and
$L$-preserving trinification theory considered here, the
electroweak anomaly with all three generations instead gives
$\Delta B=\Delta L\in3\mathbb Z$.}
 To make the symmetry argument explicit, consider the usual matter
representations 
\begin{equation}
\begin{gathered}
Q_L\sim(\mathbf3,\overline{\mathbf3},\mathbf1)_L,
\qquad
Q_R\sim(\mathbf3,\mathbf1,\overline{\mathbf3})_R,\\
\Psi_L\sim(\mathbf1,\mathbf3,\overline{\mathbf3})_L,
\end{gathered}
\end{equation}
together with three bifundamental Higgs fields
$\Phi_{1,2,3}\sim(\mathbf1,\mathbf3,\overline{\mathbf3})$.
The VEVs above belong to $\Phi_1$ and $\Phi_2$, while $\Phi_3$
supplies a second independent Yukawa matrix \cite{Raut:2022}.

Define $U(1)_F$ by
\begin{equation}
\begin{gathered}
F(Q_L)=F(Q_R)=\frac23,
\qquad
F(\Psi_L)=0,\\
F(\Phi_1)=F(\Phi_3)=0,
\qquad
F(\Phi_2)=1,
\end{gathered}
\end{equation}
and let
\begin{equation}
T_L=T_R=\frac13\operatorname{diag}(1,1,-2),
\qquad
X=F+T,
\end{equation}
where $T=T_L+T_R$. The generator $X$ acts on the Standard Model fields as
\begin{equation}
\begin{aligned}
X(Q)&=X(u_R)=X(d_R)=\frac13,\\
X(\ell)&=X(e_R)=X(\nu_R)=1,
\end{aligned}
\end{equation}
and therefore restricts to $B+L$ on the Standard Model fermions. The exotic quarks $D_L$ and $D_R$
have $X=4/3$, while the exotic leptons are $X$-neutral.
The additional neutral singlet $N=(\Psi_L)_{33}$ has $X=0$
and may acquire a symmetry-preserving mass, whereas
$X(\nu_R)=1$ forbids a Majorana mass and leaves the light
neutrinos Dirac. Both
high-scale VEVs and electroweak VEVs in the upper $2\times2$ blocks
of $\Phi_1$ and $\Phi_3$ preserve $X$.

Assigning
\begin{equation}
B(Q_L)=B(Q_R)=\frac13,
\qquad
B(\Psi_L)=B(\Phi_a)=0,
\end{equation}
and taking the scalar potential to preserve $U(1)_F$, the Yukawa
operators
\begin{equation}
Q_LQ_R^c\Phi_{1,3},
\qquad
(\Psi_L\Psi_L\Phi_{1,3})_{\epsilon\epsilon}
\end{equation}
preserve both $B$ and $X$. The combination
\begin{equation}
B-L=2B-X
\end{equation}
is therefore also preserved. Thus $X$ and $B-L$ are classical symmetries of the full trinified
theory, with $X$ reducing to $B+L$ on the Standard Model fields.

Quantum mechanically, however, $X$ has the mixed anomalies
\begin{equation}
\begin{aligned}
\mathcal A_{U(1)_X SU(3)_C^2}&=0,\\
\mathcal A_{U(1)_X SU(3)_L^2}&=N_g,\\
\mathcal A_{U(1)_X SU(3)_R^2}&=-N_g.
\end{aligned}
\end{equation}
Thus the exotic fermions do not complete $B+L$ to an anomaly-free
continuous symmetry. For either global form, the anomaly preserves
$\mathbb Z_{2N_g}\subset U(1)_X$.

The elementary trinification monopole has $(\alpha,\beta,\gamma)=(3,1,0)$ and activates the bulk 
anomaly; the contractions derived in Section~\ref{sec:sm-monopoles}  give the radial anomalies governing fermion scattering in the monopole background,
\begin{equation}
\mathcal A_X^{(Y)}=-3N_g,
\qquad
\mathcal A_X^{(I)}=N_g.
\end{equation}
Their nonzero values forbid an ordinary gauge-invariant core boundary condition that preserves continuous $U(1)_X$. 

Combining the exact $\mathbb Z_{2N_g}$ subgroup with $B-L$ conservation gives the selection rule 
\begin{equation}
\Delta B=\Delta L\in N_g\mathbb Z.
\end{equation}
For $N_g=3$, the minimal nonzero channel has the expected $|\Delta B|=|\Delta L|=3.$

Unlike the Pati--Salam monopole, the trinification monopole admits no spherically symmetric $SU(2)$ embedding. The Wilkinson--Goldhaber
criterion requires
\begin{equation}
Q_{\rm mg}=2(I_3-T_3),
\end{equation}
where $I_a$ and $T_a$ generate $SU(2)$ subgroups of the unbroken and
full gauge groups, respectively \cite{Wilkinsin:1977}. The projection
of $I_a$ onto $\mathfrak{su}(3)_R$ vanishes, so the $SU(3)_R$
component of $T_3$ would have to be
\begin{equation}
(T_3)_R
=-\frac12(Q_{\rm mg}^{(3)})_R
=\operatorname{diag}\left(1,-\frac12,-\frac12\right).
\end{equation}
Under any $SU(2)\subset SU(3)$, the fundamental branches as
$\mathbf3$, $\mathbf2\oplus\mathbf1$, or
$\mathbf1\oplus\mathbf1\oplus\mathbf1$, giving $T_3$ eigenvalues
\begin{equation}
(1,0,-1),
\qquad
\left(\frac12,-\frac12,0\right),
\qquad
(0,0,0),
\end{equation}
up to permutation. None agrees with the required spectrum. The
monopole is therefore necessarily nonspherical \cite{Lazarides:1989}. 

This prevents a Sen-style calculation of the condensate and scattering rate, but does not alter the anomaly selection rule. Moreover, we can assemble various other pieces of evidence for the expected $B+L$ violation in monopole-fermion scattering. First, spherical symmetry is not required to count the protected radial
fermion modes. On a large sphere surrounding the monopole, a fermion
component of weight $w$ couples to a line bundle of degree
\begin{equation}
q_w=\frac{1}{2\pi}\int_{S^2}w(F)=w(Q_{\rm mg}).
\end{equation}
The angular Dirac operator therefore obeys
\begin{equation}
\operatorname{ind}D_{S^2,w}
=n_+-n_-=q_w.
\end{equation}
Its $|q_w|$ protected angular zero modes become chiral radial channels
in the effective two-dimensional scattering problem. Their
wavefunctions may be distorted and mixed by the nonspherical
monopole background, but their net chirality, and hence the radial
anomaly they carry, is fixed by the asymptotic magnetic charge.

Second, the four-dimensional index theorem identifies a
light-fermion operator with the charges required by the selection
rule. To relate this operator to the monopole flux, consider the
theory on $S^2\times S^2$. Put the monopole flux through the first
sphere and one unit of flux in an unbroken weak-isospin direction
through the second. In the two trinification factors, the corresponding
flux generators are
\begin{equation}
\begin{aligned}
m_L&=\operatorname{diag}(0,-1,1),
&
m_R&=\operatorname{diag}(-2,1,1),\\
h_L&=\operatorname{diag}(1,-1,0),
&
h_R&=0.
\end{aligned}
\end{equation}
The generators $h_i$ commute with the magnetic charge, preserve both
high-scale Higgs VEVs, and satisfy $\exp(2\pi i h_i)=1$. The second
Chern numbers of this background are therefore fixed by the two
fluxes \cite{Reece:2025}:
\begin{equation}
\begin{aligned}
k_i
&=\frac{1}{8\pi^2}\int_{S^2\times S^2}
  \operatorname{tr}(F_i\wedge F_i)
=\operatorname{tr}(m_i h_i),\\
(k_L,k_R)&=(1,0).
\end{aligned}
\end{equation}
Thus the configuration carries one unit of $SU(3)_L$ instanton
number and none under $SU(3)_R$. The integrated mixed anomaly gives
\begin{equation}
\Delta X
=2\left(
\mathcal A_{U(1)_XSU(3)_L^2}k_L
+\mathcal A_{U(1)_XSU(3)_R^2}k_R
\right)
=2N_g.
\end{equation}

The four-dimensional indices are three for $Q_L$ and three for
$\Psi_L$ per generation. These count net chiralities, not all
zero modes. The VEV $\langle\Phi_1\rangle=v_1E_{33}$ gives mass
to the exotic doublets with $SU(3)_R$ indices $1$ and $2$.
For full-rank exotic mass matrices, the net light-fermion indices
are three quarks and one lepton per generation. A lowest-dimensional
gauge-invariant operator with these chiral charges is
\begin{equation}
\begin{aligned}
\mathcal O_{\rm tri}
&\sim\prod_{a=1}^{N_g}(QQQ\ell)_a.
\end{aligned}
\end{equation}
This gives an explicit light-fermion operator with
$\Delta B=\Delta L=N_g$.

Finally, we can ask whether nonsphericity introduces new parametric suppression in the scattering rate. Sen's explicit calculation used a spherical model, but
his argument for its broader applicability assumes only that the
monopole core can be represented by a boundary condition and that,
outside the core, the field approaches the monopole field fixed by
the total magnetic flux \cite{Sen-1985-II}. A localized nonspherical
monopole has the same leading $1/r^2$ field, while its nonspherical
multipoles fall more rapidly. Nonsphericity may therefore change
angular overlaps and numerical coefficients, but it supplies no new
small parameter. Sen's estimate consequently gives no power suppression by
the core size, weak coupling, or electroweak scale provided Eq.~(\ref{eq:unsuppressed}) holds. For a GUT-scale core, this condition is easily satisfied even at
hadronic energies.

The elementary trinification monopole therefore provides a second
setting in which a classically $B+L$- and $B-L$-preserving completion
of the Standard Model must violate $B+L$ in monopole-fermion scattering, with
$|\Delta B|=|\Delta L|=3$ as the minimal nonzero channel.
Nonsphericity introduces no new parametric suppression in this
argument, although an explicit nonspherical solution is still needed
to determine the detailed scattering amplitudes.

\section{Conclusions}

Generalized symmetries organize the fate of Standard Model flavor symmetries in the presence of magnetic monopoles. As we have seen, in a monopole background the relevant four-dimensional mixed anomalies contract with the magnetic charge and become anomalies of the radial scattering problem. The resulting obstruction to a symmetry-preserving boundary condition realizes the emergence constraints implied by the generalized symmetry structure and provides a direct mechanism for flavor violation in monopole--fermion scattering.

For $U(1)_{B+L}$, the electroweak anomaly preserves $\mathbb Z_{2N_g}$, which together with $B-L$ conservation restricts any violation to $\Delta B=\Delta L\in N_g\mathbb Z.$ We have recovered this result explicitly for the Pati--Salam monopole studied by Sen and found the same selection rule for the elementary trinification monopole, where the radial anomaly and index theorems identify a light-fermion operator with the required charges despite the absence of spherical symmetry. Both these concrete examples and more general arguments support the expectation that anomaly-induced $B+L$ violation is a generic feature of monopoles whose microscopic magnetic fields activate the electroweak anomaly.

Although the electroweak anomaly fixes the obstruction and selection rule, it does not determine the scattering rate. Sen's argument indicates that the effect is not power suppressed by the monopole size, weak coupling, or electroweak scale when $r_0<g^2E/m_W^2$, and we find no new parametric suppression associated with monopole nonsphericity. This leads to the compelling possibility of highly distinctive and experimentally observable $\Delta B = \Delta L = 3$ processes catalyzed by monopoles in terrestrial or astrophysical settings. Our work leaves a number of open directions, including a quantitative treatment of the nonspherical trinification monopole and its scattering problem, as well as the detailed phenomenology of $\Delta B=\Delta L=3$ monopole catalysis.

\acknowledgments

We would like to thank Daniel Brennan, Jacques Distler, Vivian Gao, Howard Georgi, Jake Hauser, Marius Kongsore, and Seth Koren for useful conversations. We would also like to thank Marius Kongsore for helpful comments on the draft. ChatGPT-5.6 was used to perform adversarial audits of results in the manuscript. This work was supported in part by the U.S. Department of Energy under the grant DE-SC0011702. This work was performed in part at the Kavli Institute for Theoretical Physics, supported by the National Science Foundation under the grant NSF PHY-2309135.

\appendix
\section{Notation and Conventions}

\begin{table}[!htbp]
\centering
\begin{tabular}{|c||c|c|c|c|c|}
\hline & $SU(3)_C$ & $SU(2)_L$ & $U(1)_Y$ & $U(1)_B$ & $U(1)_L$ \\
\hline \hline $Q=\binom{u}{d}$ & $\mathbf{3}$ & $\mathbf{2}$ & +1/6 & +1/3 & 0 \\
\hline $u_R$ & $\mathbf{3}$ & $\mathbf{1}$ & +2/3 & +1/3 & 0 \\
\hline $d_R$ & $\mathbf{3}$ & $\mathbf{1}$ & $-1/3$ & +1/3 & 0 \\
\hline $\ell=\binom{\nu}{e}$ & $\mathbf{1}$ & $\mathbf{2}$ & $-1/2$ & 0 & +1 \\
\hline $e_R$ & $\mathbf{1}$ & $\mathbf{1}$ & $-1$ & 0 & +1 \\
\hline $\nu_R$ & $\mathbf{1}$ & $\mathbf{1}$ & 0 & 0 & +1 \\
\hline
\end{tabular}
\caption{Charges of Weyl fermions in the Standard Model. We define $B=\frac{1}{3}(Q+u_R+d_R)$. In the $U(N_g)^5$ flavor analysis, $L=\ell+e_R$. In the Pati--Salam and trinification discussions, $\nu_R$ is included and $L=\ell+e_R+\nu_R$.}
\label{tab:sm-fermions}
\end{table}

Throughout this paper, $Q=(u,d)^T$ and $\ell=(\nu,e)^T$ denote left-handed Weyl fermions, while $u_R$, $d_R$, $e_R$, and $\nu_R$ denote right-handed Weyl fermions. Their Standard Model gauge, baryon, and lepton charges are listed in Table~\ref{tab:sm-fermions}. A Weyl fermion enters a chiral anomaly coefficient with the relative sign $(-1)^{\chi_f}$, where $\chi_f=0$ for a left-handed fermion and $\chi_f=1$ for a right-handed fermion. When a unified representation is most naturally displayed using charge-conjugated fields, every such field will be labeled explicitly as $f_R^c$. The hypercharge normalization in Table~\ref{tab:sm-fermions} is conventional, and we write $\widehat Y=6Y$ when integer-normalized hypercharge is required.

In the absence of Yukawa couplings, the Standard Model enjoys a $U(N_g)^5/U(1)_Y$ 0-form global symmetry, with each $U(N_g)_f$ ($f=Q,u,d,\ell,e$) factor corresponding to rotations of the $N_g$ generations of each fermion type into each other. Our conventions are such that the $N_g$ generations of each Standard Model Weyl fermion in Table \ref{tab:sm-fermions} transform together in the fundamental representation of their corresponding $U(N_g)_f$.

We use the conventional $SU(N)$ generator normalization $T(\mathbf{N})=\frac{1}{2}$. For Abelian generators $X,Y$ and a non-Abelian gauge factor $G$, we define
\begin{align}
\mathcal A_{XG^2}
    &=\sum_f(-1)^{\chi_f}X_fT_G(R_f),\\
\mathcal A_{X^2Y}
    &=\sum_f(-1)^{\chi_f}X_f^2Y_f,\\
\mathcal A_{XY^2}
    &=\sum_f(-1)^{\chi_f}X_fY_f^2.
\end{align}
A unit $G$ instanton violates $X$ by
\begin{equation}
\Delta X=2\mathcal A_{XG^2}.
\end{equation}
Coefficients of the form $\mathcal A_{F^2Y}$ govern flavor-flavor-hypercharge 't Hooft anomalies, while $\mathcal A_{XY^2}$ governs the ABJ anomaly of the $X$ current. For integer-normalized hypercharge $\widehat Y=6Y$,
\begin{equation}
\mathcal A_{X^2\widehat Y}=6\mathcal A_{X^2Y},
\qquad
\mathcal A_{X\widehat Y^2}=36\mathcal A_{XY^2}.
\end{equation}

\bibliography{apssamp}% Produces the bibliography via BibTeX.

\end{document}